\documentclass[journal, 11pt, draftcls, onecolumn]{IEEEtran}

\usepackage{amsmath}
\usepackage{amssymb}
\usepackage{times}
\usepackage{color}
\usepackage{graphicx}
\usepackage{cite,url}
\usepackage{pst-all}

\newcommand{\insertfig}[4]{
\begin{figure}[ht]
\centerline{\includegraphics[width=#1\columnwidth]{#2.eps}}
\caption{#3}\label{#4}\end{figure}}

\DeclareMathAlphabet{\mathsfbf}{OT1}{cmss}{sbc}{n}

\newcommand{\example}[2]{
\begin{center}
\parbox{0.9\columnwidth}{
\rule{0.9\columnwidth}{0.5mm}\\
\noindent {\bf Example~#1:}
#2 \\
\rule{0.9\columnwidth}{0.5mm}
}
\end{center}
}

\newtheorem{theorem}{Theorem}[section]
\newtheorem{lemma}{Lemma}[section]

\newcommand{\EE}{\mathop{\mathbb{E}}\limits} 
\newcommand{\RR}{\mathbb{R}} 
\newcommand{\Herm}{^\dagger} 
\newcommand{\Tran}{^{\rm T}} 

\newcommand{\ee}{{\rm e}}
\newcommand{\jj}{{\rm j}}  
\newcommand{\dd}{{\rm\,d}} 

\newcommand{\av}{{\bf a}}

\newcommand{\nv}{{\bf n}}

\newcommand{\pv}{{\bf p}}
\newcommand{\qv}{{\bf q}}

\newcommand{\sv}{{\bf s}}

\newcommand{\wv}{{\bf w}}
\newcommand{\vv}{{\bf v}}
\newcommand{\xv}{{\bf x}}
\newcommand{\yv}{{\bf y}}

\newcommand{\zerov}{{\bf 0}}

\newcommand{\ellv}{\boldsymbol{\ell}}

\newcommand{\Gm}{{\bf G}}

\newcommand{\Lm}{{\bf L}}

\newcommand{\Qm}{{\bf Q}}

\newcommand{\Tm}{{\bf T}}

\newcommand{\Hc}{{\cal H}}

\newcommand{\Lc}{{\cal L}}
\newcommand{\Mc}{{\cal M}}

\newcommand{\Pc}{{\cal P}}
\newcommand{\Qc}{{\cal Q}}

\newcommand{\Sc}{{\cal S}}

\newcommand{\Xc}{{\cal X}}

\newcommand{\gammav}{\boldsymbol{\gamma}}

\newcommand{\muv}{\boldsymbol{\mu}}

\newcommand{\omegav}{\boldsymbol{\omega}}

\def\trace{\mathsf{Tr}}

\newcommand{\eqdef}{\ensuremath{\stackrel{\mbox{\upshape\tiny $\Delta$}}{=}}}
\def\ben{\begin{enumerate}}
\def\beq{\begin{equation}}
\def\beqa{\begin{eqnarray}}
\def\bit{\begin{itemize}}
\def\een{\end{enumerate}}
\def\eeq{\end{equation}}
\def\eeqa{\end{eqnarray}}
\def\eit{\end{itemize}}

\def\non{\nonumber\\}

\def\union{\mathop{\cup}\limits}

\def\limbeta{\lim_{\substack{M,r\rightarrow +\infty \\\beta}}}

\title{On the $d$-dimensional Quasi-Equally\\ Spaced Sampling
}

\author{Alessandro Nordio$^\star$, Carla-Fabiana Chiasserini$^\star$, Emanuele Viterbo$^\ddag$
\vspace{3mm}\\
$^\star$ Dipartimento di Elettronica, Politecnico di Torino\\
        C. Duca degli Abruzzi 24, I-10129 Torino, Italy\\
        Phone: +39 011 090 4226; Fax +39 011 0904099 \\
E-mail: {\{alessandro.nordio,chiasserini\}@polito.it}
\vspace{3mm}\\
$^\ddag$ DEIS, Universit\`a della Calabria\\
Via P. Bucci, Cubo 42C, 87036 Rende (CS), Italy\\
        Phone: +39 0984 494778; Fax +39 0984 494713 \\
E-mail: viterbo@deis.unical.it
}

\begin{document}
\maketitle

\begin{abstract}
We study a class of random matrices that appear in
several communication and signal processing applications, and whose
asymptotic eigenvalue distribution is closely related to the
reconstruction error of an irregularly sampled bandlimited signal.
We focus on the case where the random variables characterizing these
matrices are $d$-dimensional vectors, independent, and quasi-equally
spaced, i.e., they have an arbitrary distribution and their averages
are vertices of a $d$-dimensional grid. Although a closed form
expression of the eigenvalue distribution is still unknown, under
these conditions we are able ({\em i}) to derive the distribution
moments as the matrix size grows to infinity, while its aspect ratio
is kept constant, and ({\em ii}) to show that the eigenvalue
distribution tends to the Mar\v{c}enko-Pastur law as $d \rightarrow
\infty$. These results can find application in several fields, as an
example we show how they can be used for the estimation of the mean
square error provided by linear reconstruction techniques.
\end{abstract}

{\bf EDICS:} DSP-RECO Signal reconstruction, DSP-SAMP Sampling,
SPC-PERF Performance analysis and bounds.

\section{Introduction}

Consider the class of random matrices of size $(2M+1) \times r$,
with entries given by
\begin{equation}
 \Gm = \frac{1}{\sqrt{2M+1}}\left[
  \begin{array}{ccc}
  \ee^{-\jj 2 \pi M x_1} & \cdots &\ee^{-\jj 2 \pi M x_r} \\
  \vdots & & \vdots \\
  1  & \cdots & 1\\
  \vdots & & \vdots \\
  \ee^{+\jj 2 \pi M x_1} & \cdots &\ee^{+\jj 2 \pi M x_r}
  \end{array}
  \right]
\label{eq:Gkq}
\end{equation}
The generic element of $\Gm$ can be written as:
$\Gm_{\ell,q}=\frac{1}{\sqrt{2M+1}} \ee^{\jj 2 \pi \ell x_q}$, $\ell =
-M, \ldots, M$, $q=0,\ldots, r-1$, where $x_q$ are independent random
variables characterized by a probability density function (pdf)
$f_{x_q}(z)$, with $0 \le z \le 1$.
These matrices are  Vandermonde matrices with complex exponential entries;
 they appear in many signal/image processing applications and
  have been studied in a number of recent works,
  (see e.g.,~\cite{infocom07,TSP1_08,ipsn07,izs08,TSP2_08,icc08,RyanDebbah1,RyanDebbah2}).
More specifically, in the field of signal processing 
for sensor networks, \cite{infocom07,TSP1_08} studied the
performance of linear reconstruction techniques for physical fields
irregularly sampled by sensors. In such scenario, the random
variables $x_q$ in~(\ref{eq:Gkq}) represent the coordinates of the
sensor nodes. The work in~\cite{ipsn07} addressed the case where
these coordinates are uniformly distributed and subject to an
unknown jitter.
In the field of communications, 
the study in~\cite{RyanDebbah2} presented a number of applications
where these matrices appear, which range
from multiuser MIMO systems to multifold scattering.

In spite of their numerous applications,
few results are known for the Vandermonde matrices in
(\ref{eq:Gkq}). In particular, a closed form expression
  for the eigenvalue distribution of the Hermitian Toeplitz matrix
  $\Gm\Gm\Herm$, as well as its asymptotic behavior, would be of great
  interest. As an example, in \cite{infocom07,TSP1_08,icc08}, it has been observed
  that the performance of linear techniques for reconstructing a
  signal from a set of irregularly-spaced samples with known
  coordinates is a function of the {\em asymptotic eigenvalue
    distribution} of $\Gm\Gm\Herm$.
  The asymptotic eigenvalue distribution of $\Gm\Gm\Herm$ is defined as the
  distribution of its eigenvalues, in the limit
  of $M$ and $r$ growing to infinity while their ratio
is kept constant.  Unfortunately, such distribution is still unknown.

In this work, we consider a general formulation which extends the
model in $(\ref{eq:Gkq})$ to the $d$-dimensional domain. We
study the properties of random matrices of size $(2M+1)^d\times r$ and
entries given by
\begin{equation}
 (\Gm_d)_{\nu(\ellv), q} = \frac{1}{\sqrt{(2M+1)^d}}\ee^{-\jj 2
    \pi\ellv\Tran \xv_q}
 \label{eq:Gd}
\end{equation}
where the vectors $\xv_q=[x_{q1},\ldots, x_{qd}]\Tran$ have
independent entries, characterized by the pdf $f_{x_{qm}}(z)$,
$q=0,\ldots,r-1$, $m=1,\ldots,d$, and $d$ is the number
  of dimensions.  The invertible function
\begin{equation}
\nu(\ellv) = \sum_{m=1}^d (2M+1)^{m-1} \ell_m
\label{eq:nu}
\end{equation}
maps the vector of integers $\ellv = [\ell_1, \ldots, \ell_d]\Tran$,
$\ell_m=-M,\ldots,M$ onto a scalar index, i.e., the row index of the
matrix $\Gm_d$.
Notice that, when $d=1$, $\Gm_d$ reduces to~(\ref{eq:Gkq}).

For the matrix model in~(\ref{eq:Gd}), we study the interesting
case where $\xv_q$ are independent, {\em quasi-equally spaced} random
variables in the $d$-dimensional hypercube $[0,1)^d$. In other words,
  we assume that the averages of $\xv_q$ are the vertices of a
  $d$-dimensional grid in $[0,1)^d$.  This is often the case arising
    in measurement systems affected by jitter, or in sensor network
    deployments where the sensors sampling the physical field can
    only be roughly placed at equally spaced positions, due to
    terrain conditions and deployment practicality \cite{Ganesan}.
    Note that the distribution of the random variables $\xv$ can be of
    any kind, the only assumption we make is on their averages being
    equally spaced.
Since an analytic expression of the eigenvalue distribution
of $\Gm_d\Gm_d\Herm$ is unknown,
we derive a closed form    expression for its moments.
This enables us to show that, as $d
    \rightarrow \infty$, the eigenvalue distribution tends to the
    Mar\v{c}enko-Pastur law~\cite{MarcenkoPastur}.  At the
    end of the paper, we present some numerical results and
    applications where the moments and the
    asymptotic approximation to the eigenvalue distribution
of $\Gm_d\Gm_d\Herm$ can be of great use.

\section{Previous Results and Problem Formulation}
As a first step, we briefly review previous results on the $\Gm_d$
matrices.  In a one-dimensional domain ($d=1$), the work
in~\cite{infocom07} considered an irregularly sampled bandlimited
signal, which is reconstructed  using linear techniques and
assuming the samples coordinates to be known.  The performance of the
reconstruction system was derived as a function of the eigenvalue
distribution $f_\lambda(1,\beta,z)$ of the matrix $\Tm_1
=\beta\Gm_1\Gm_1\Herm$, where $\beta$ is the aspect ratio%
\footnote{The aspect ratio of $\Gm$ is the ratio between
the number of rows and the number of columns of the matrix} 
of $\Gm_1$~\cite{infocom07,TSP1_08}.  An explicit expression of
the moments
\[ \EE[\lambda_{1,\beta}^p] = \int_0^\infty z^p f_\lambda(1,\beta,z) \dd z \]
was attained in~\cite{izs08,TSP2_08}, for the specific case where $x_q$ are
uniformly distributed in $[0,1)$. Also, in the case where $x_q$ are
  independent, quasi-equally spaced random variables, the analytic
  expression of the second moment of the eigenvalue distribution of
  $\Tm$, i.e., $\EE[\lambda_{1,\beta}^2]$, was obtained
  in~\cite{ipsn07}.  Then, in~\cite{RyanDebbah1} the moments
  $f_\lambda(1,\beta,z)$ were derived for an arbitrary distribution
  $f_{x_q}(z)$.

In~\cite{izs08,TSP2_08}, the $d$-dimensional model~(\ref{eq:Gd})
  was also investigated.  There, the properties of the random matrices
  $\Gm_d$ were studied in the case where the vectors
  $\xv_q=[x_{q1},\ldots, x_{qd}]\Tran$ have independent entries, {\em
    uniformly distributed} in the hypercube $[0,1)^d$.  Under such
    assumptions, and for given $d$ and aspect ratio $\beta$,
an analytic expression
    of the moments of $f_\lambda(d,\beta,z)$ was derived and it was
    shown that, as $d\rightarrow \infty$, $f_\lambda(d,\beta,z)$ tends
    to the Mar\v{c}enko-Pastur  law~\cite{MarcenkoPastur}, i.e.,
\[
\lim_{d \rightarrow \infty} f_\lambda(d,\beta,z) = f_{\rm MP}(\beta,z) =
\frac{\sqrt{(c_1-z)(z-c_2)}}{2\pi z \beta}
\]
where $c_1,c_2 = (1 \pm \sqrt{\beta})^2$, $0 < \beta \le 1$, $c_2 \le
x \le c_1$.

The following sections detail the problem addressed in this work and introduce some
useful notations.

\subsection{The quasi-equally spaced multidimensional model}
\label{sec:qes_multi}

We consider the matrix class
  in~(\ref{eq:Gd}) and assume that the vectors $\xv$ are independent,
quasi-equally spaced random variables in the $d$-dimensional hypercube
$[0,1)^d$, i.e., the averages of $\xv$ are the vertices of a
  $d$-dimensional grid in $[0,1)^d$.

We define $\rho$ as the number of vertices per dimension, thus the
total number of vertices is $r=\rho^d$.
We denote the coordinate of a generic vertex of the grid by the
vector $\qv/\rho \in [0,1)^d$, where $\qv=[q_1,\ldots, q_d]\Tran$,
is an integer vector and
  $q_m=0,\ldots, \rho-1$.  For notation simplicity and in analogy
  with~(\ref{eq:nu}), we identify the vertex with coordinate
  $\qv/\rho$ by the scalar index
\begin{equation}
\mu(\qv) = \sum_{m=1}^d \rho^{m-1} q_m
\label{eq:mu}
\end{equation}
Notice that $0 \le \mu(\qv) \le r-1$ is an invertible function and
allows us to write
\[ \xv_{\mu(\qv)} = \frac{\qv}{\rho} + \frac{\tilde{\xv}_{\mu(\qv)}}{\rho} \]
where the average
\[ \EE[\xv_{\mu(\qv)}] = \frac{\qv}{\rho} + \frac{\mathbf 1}{2\rho} \]
is the coordinate of the sample identified by the scalar label
$\mu(\qv)$ and {\bf 1} is the all ones vector.  Furthermore, we
assume that the entries of the vectors $\tilde{\xv}_{\mu(\qv)}$ are
i.i.d. with pdf $f_{\tilde{x}}(z)$ which does not depend on $r$,
$M$, or $\qv$. By using this notation, the entries of $\Gm_d$ are
then given by
\begin{equation}
 (\Gm_d)_{\nu(\ellv), \mu(\qv)} = \frac{1}{\sqrt{(2M+1)^d}}\ee^{-\jj 2
    \pi   \ellv\Tran \xv_{\mu(\qv)}}
\end{equation}
while the aspect ratio is
\begin{equation}
\beta = \frac{(2M+1)^d}{r} = \left(\frac{2M+1}{\rho}\right)^d
\label{eq:beta_qes}
\end{equation}
The Hermitian Toeplitz matrix $\Tm_d =
\beta\Gm_d\Gm_d\Herm$ is defined as
\begin{equation}
(\Tm_d)_{\nu(\ellv),\nu(\ellv')} = \frac{1}{\rho^d}\sum_{\qv}
\ee^{-\jj 2 \pi \xv_{\mu(\qv)}\Tran(\ellv-\ellv') }
\label{eq:t_d}
\end{equation}
where $\sum_{\qv}$ represents a $d$-dimensional sum over all vectors $\qv$ such
that $q_m=0,\ldots,\rho-1$, $m=1,\ldots,d$.

Our goals are {\em (i)} to derive the analytic expression of the
moments of $f_\lambda(d,\beta,z)$ with quasi-equally spaced vectors
$\xv_{\mu(\qv)}$ (Section~\ref{sec:moments}), and {\em (ii)} to show
that as $d\rightarrow \infty$, $f_\lambda(d,\beta,z)$ tends to the
Mar\v{c}enko-Pastur law (Section~\ref{sec:convergence}).

\section{Closed form expression of the moments of the asymptotic eigenvalue pdf}
\label{sec:moments}
Following the approach adopted in~\cite{Billingsley,Tulino}, in the limit
for $M$ and $r$ growing to infinity with constant aspect ratio $\beta$
and dimension $d$, we compute the closed form expression of
$\EE[\lambda_{d,\beta}^p]$, which can be obtained from the powers of
$\Tm_d$ as~\cite{TulinoVerdu},
\begin{equation}
\EE[\lambda_{d,\beta}^p]
 = \limbeta \frac{\trace\big\{\EE_{\Xc}\left[\Tm_d^p\right]\big\}}{(2M+1)^d}
\label{eq:lambda_p}
\end{equation}
In~(\ref{eq:lambda_p}) the symbol $\trace$ identifies the matrix trace
operator, and the average $\EE_{\Xc}[\cdot]$ is computed over the set
of random variables $\Xc=\{\xv_0, \ldots, \xv_{r-1}\}$. An efficient
method to compute~(\ref{eq:lambda_p}) exploits {\em set partitioning}.
Indeed, note that the power $\Tm_d^p$ is the matrix product of $p$
copies of $\Tm_d$.  This operation yields exponential terms, whose
exponents are given by a sum of $p$ terms of the form
$\xv_{\mu(\qv_i)}\Tran(\ellv_i-\ellv_{[i+1]})$ (see also
(\ref{eq:lambda_p2}) in Appendix~\ref{app:proof_theorem1}).  The
average of this sum depends on the number of distinct vectors $\qv_i$, and
all possible cases can be described as partitions of the set
$\Pc=\{1,\ldots,p\}$.  In particular, the case where in the set
$\{\qv_1, \ldots, \qv_p\}$ there are $1\le k \le p$ distinct vectors,
corresponds to a partition of $\Pc$ in $k$ subsets. It follows that
a fundamental step to calculate (\ref{eq:lambda_p}) is the computation
of all possible partitions of set $\Pc$. Before
proceeding further in our analysis, we therefore introduce some useful
definitions related to set partitioning.

\subsection{Definitions}
\label{sec:definitions}
Let the integer $p$ denote the moment order and let the vector
$\muv=[\mu_1,\ldots,\mu_p]$ be a possible combination of $p$
integers. In our specific case, each entry of the vector $\muv$ is
given by the expression in (\ref{eq:mu}), i.e., $\mu_i=\mu(\qv_i)$ and,
thus, can range between $0$ and $r-1$.

We define:
\begin{itemize}
\item the scalar integer $1 \le k(\muv)\le p$ as the number of
  distinct entries
  of the vector $\muv$;
\item $\gammav(\muv)$ as the vector of integers, of length $k(\muv)$,
  whose entries $\gamma_j(\muv)$, $j=1,\ldots, k(\muv)$, are the
  entries of $\muv$ without repetitions, in order of appearance within $\muv$;
\item $\Pc_j(\muv)$ as the set of indices of the entries of $\muv$
  with value $\gamma_j(\muv)$, $j=1,\ldots,k(\muv)$;
\item the vector $\omegav(\muv)=[\omega_1(\muv), \ldots,
  \omega_p(\muv)]$ such that, for any given $j=1,\ldots, k(\muv)$, we
  have $\omega_i(\muv) = j$ if $i \in \Pc_j(\muv)$, $i=1,\ldots,p$.

\medskip

\example{1}{ Let $\muv = [1,5,2,8,5,3,2]$, then $k(\muv) = 5$ since
  the entries of $\muv$ take 5 distinct values (i.e.,
  $\{1,5,2,8,3\}$).  Such values, taken in order of
    appearance in $\muv$ form the vector $\gammav(\muv) =
    [1,5,2,8,3]$. The value $\gamma_1=1$ appears at position $1$ in
    $\muv$, therefore $\Pc_1(\muv) = \{1\}$.  The value $\gamma_2=5$
    appears at positions 2 and 5 in $\muv$, therefore $\Pc_2(\muv) =
    \{2,5\}$. Similarly $\Pc_3(\muv) = \{3,7\}$, $\Pc_4(\muv)=\{4\}$,
    and $\Pc_5(\muv) = \{6\}$. By using the sets $\Pc_j$ we build the
    vector, $\omegav(\muv)$. For each $j=1,\ldots,k$ we assign the
    value $j$ to every $\omega_i$ such that $i\in \Pc_j$. For example,
    $\omega_2=\omega_5=2$ since the integers 2 and 5 are in
    $\Pc_2$. In conclusion $\omegav(\muv)= [1,2,3,4,2,5,3]$.}

\medskip

Furthermore, we define:
\item $\Omega_p$ as the set of partitions of $\Pc$;
\item $\Omega_{p,k}$ as the set of partitions of $\Pc$ in $k$ subsets,
$1\le k \le p$, with $\union_{k=1}^p \Omega_{p,k} = \Omega_p$.
\end{itemize}
Note that: {\em (i)} the cardinality of $\Omega_p$, denoted by
$B(p)=|\Omega_p|$, is the $p$-th Bell number~\cite{bellnumbers} and
{\em (ii)} the cardinality of $\Omega_{p,k}$, denoted by
$S(p,k)=|\Omega_{p,k}|$, is a Stirling number of the second
kind~\cite{stirling2numbers}.

From the above definitions, it follows that:
\begin{enumerate}
  \item the vector $\muv$ induces a partition of the set $\Pc$ which
    is identified by the subsets $\Pc_j(\muv)$. These subsets have the
    following properties
    \[ \union_{j=1}^{k(\muv)} \Pc_j(\muv) =
    \Pc, \hspace{1cm}\Pc_j(\muv) \cap \Pc_{j'}(\muv) = \emptyset \quad
    \mbox{for} j\neq j'\] Even though the partition identified by
    $\muv$ is often represented as $\{\Pc_1, \ldots, \Pc_{k(\muv)}\}$,
    by its definition, an equivalent representation of such partition
    is given by the vector $\omegav(\muv)$. Therefore, from now on we
    will refer to $\omegav(\muv)$ as a partition of the $p$ element
    set $\Pc$ induced by $\muv$  (for simplicity,
      however, often we will not explicit the dependency of $\omegav$
      on $\muv$);
  \item $k(\omegav) = k(\muv)$, since the entries of $\omegav$ take
    all possible values in the set $\{1,\ldots,k(\muv)\}$;
  \item $\Pc_j(\omegav) = \Pc_j(\muv)$, for $j=1,\ldots, k(\muv)$.
\end{enumerate}

At last, we define $\Mc(\omegav)$ as the set of $\muv$ inducing the
same partition $\omegav$ of $\Pc$.

\medskip

\example{2}{Let $r=3$ and $p=3$. Since
    $\muv=[\mu_1,\ldots, \mu_p]$ and $\mu_i=0,\ldots,r-1$, $i=1,\ldots,p$, we have
    $r^p=27$ possible vectors $\muv$, namely,
    $\{[0,0,0],[0,0,1], \ldots, [2,2,1],[2,2,2]\}$.
    Each $\muv$ identifies a partition $\omegav \in \Omega_{3,k}$,
    with $k=1,\ldots, 3$, as described in Example 1.  The sets of
    partitions $\Omega_{3,k}$, are given by
    $\Omega_{3,1}=\{[1,1,1]\}$, $\Omega_{3,2} =
    \{[1,1,2],[1,2,1],[1,2,2]\}$, and $\Omega_{3,3} = \{[1,2,3]\}$,
    and have cardinality $S(3,1)=1$, $S(3,2)=3$ and $S(3,3)=1$,
    respectively.  The set of vectors $\muv$ identifying the
    partition $\omegav=[1,1,1]$, i.e., $\Mc([1,1,1])$, is given by:
    $\Mc([1,1,1]) = \{[0,0,0], [1,1,1], [2,2,2]\}$. Similarly,
    \[ \begin{array}{l}
    \Mc([1,1,2]) = \{[0,0,1], [0,0,2], [1,1,0], [1,1,2], [2,2,0],
      [2,2,1]\} \\
    \Mc([1,2,1]) = \{[0,1,0], [0,2,0], [1,0,1], [1,2,1], [2,0,2],
      [2,1,2]\} \\
    \Mc([1,2,2]) = \{[0,1,1], [0,2,2], [1,0,0], [1,2,2], [2,0,0],
      [2,1,1]\} \\
    \Mc([1,2,3]) = \{[0,1,2], [0,2,1], [1,0,2], [1,2,0], [2,0,1],
      [2,1,0]\}
    \end{array} \] }

\medskip

\subsection{Closed form expression of $\EE[\lambda_{d,\beta}^p]$}
By using the definitions in Section~\ref{sec:definitions} and by
applying set partitioning to~(\ref{eq:lambda_p}), we can state the
first main result of this work:

\begin{theorem}
\label{theorem1}
Let $\Tm_d$ be a $(2M+1)^d\times (2M+1)^d$ Hermitian random matrix as
defined in~(\ref{eq:t_d}), where the properties of the random vectors
$\xv_{\mu(\qv)}$ are described in Section~\ref{sec:qes_multi}. Then,
for any given $\beta$ and $d$, the $p$-th moment of the asymptotic
eigenvalue distribution of $\Tm_d$ is given by:
\begin{equation}
\EE[\lambda_{d,\beta}^p]
=\sum_{k=1}^p\sum_{h=1}^{k} \beta^{p-h}\sum_{\omegav \in
  \Omega_{p,k}}\sum_{\omegav' \in
  \Omega_{k,h}}u(\omegav')v(\omegav,\omegav')^d
\label{eq:theorem1}
\end{equation}
where
\begin{equation}
u(\omegav')=(-1)^{k-h}\prod_{j'=1}^h (|\Pc_{j'}(\omegav')|-1)!
\label{eq:u}
\end{equation}
\begin{equation}
v(\omegav,\omegav') = \displaystyle\left\{
 \begin{array}{ll}
  \displaystyle\int_{\Hc_p}\prod_{j=1}^{k} C\left(-\jj 2\pi\beta^{1/d}w_j(\omegav)\right)
    \dd \yv & h=1\\
  \displaystyle\int_{\Hc_p}\prod_{j=1}^{k} C\left(-\jj
  2\pi\beta^{1/d}w_j(\omegav)\right) \displaystyle\prod_{j'=1}^h \delta_D\left(\sum_{i'\in \Pc_{j'}(\omegav')}
   w_{i'}(\omegav)\right) \dd \yv & 1<h<k \\
  \displaystyle\int_{\Hc_p}\prod_{j=1}^{k}\delta_D\left( w_{j}(\omegav)\right) \dd \yv & h=k
 \end{array}
\right.
\label{eq:v(omega,omega')}
\end{equation}
and $v(\omegav,\omegav')=1$ for $k=1$.  In (\ref{eq:v(omega,omega')}),
we defined $\Hc_p$ as the $p$-dimensional hypercube $[-1/2,1/2)^p$,
  $C(s) = \EE_{\tilde{x}}[\ee^{sz}]$ as the characteristic function of
  $\tilde{x}$, $\delta_D(\cdot)$ as the Dirac's delta, and
\[w_j(\omegav) = \sum_{i \in \Pc_j(\omegav)} y_i-y_{[i+1]}\]
$y_i \in \RR$, $i=1,\ldots,p$, and $j=1,\ldots,k(\omegav)$.
\end{theorem}
\begin{IEEEproof}
The proof can be found in Appendix~\ref{app:proof_theorem1}.
\end{IEEEproof}

With the aim to give an intuitive explanation of the above
expressions, note that the right hand side of (\ref{eq:theorem1})
counts all possible partitions of the set $\Pc = \{1,\ldots,p\}$,
$C(s)$ in (\ref{eq:v(omega,omega')}) accounts for the generic
distribution of the variables $\tilde{\xv}$, and the quantity
$w_j(\omegav)$ represents the indices pairing that appears in the
exponent of the generic entry of the power $\Tm_d^p$.

To further clarify the moments computation,
Table~\ref{tab:partition_table} reports an example of partition sets
$\Omega_{n,m}$ for $n=1,\ldots,3$ and $1\leq m\leq n$, 
while Example 3 shows the
computation of the second moment of the eigenvalue distribution.

\begin{table}
\caption{Partition sets $\Omega_{n,m}$ for $n=1,2,3$, and $1\le m \le
  n$. Each partition is represented through its associated vector
  $\omegav$ and the value of $u(\omegav)$}
\begin{center}
\begin{tabular}{|c|c|c|c|} \hline
 $\omegav, u(\omegav)$  & $m=1$ & $m=2$ & $m=3$ \\ \hline
 $n=1$ & [1], 1 &     &      \\ \hline
 $n=2$ & [1,1], -1     & [1,2], 1  &       \\ \hline
 $n=3$ & [1,1,1], 2   &
    \begin{tabular}{ll}[1,1,2], & -1 \cr
                       [1,2,1], & -1 \cr
                       [1,2,2], & -1
    \end{tabular}  & [1,2,3], 1  \\ \hline
\end{tabular}
\end{center}
\label{tab:partition_table}
\end{table}

\example{3}{We compute the analytic expression of
  $\EE[\lambda_{d,\beta}^2]$.
Using~(\ref{eq:theorem1}), we get:
\[
\EE[\lambda_{d,\beta}^2]
=\sum_{k=1}^2\sum_{h=1}^{k} \beta^{2-h}\sum_{\omegav \in
  \Omega_{2,k}}\sum_{\omegav' \in
  \Omega_{k,h}}u(\omegav')v(\omegav,\omegav')^d \]
By expanding this expression and using Table~\ref{tab:partition_table}, we obtain
\[ \EE[\lambda_{d,\beta}^2] = \beta v([1,1],[1])^d - \beta
v([1,2],[1,1])^d+
v([1,2],[1,2])^d \]
We notice that, for  $k=1$, $v([1,1],[1])=1$. The term
$v([1,2],[1,2])$ refers instead to the case $k=h=2$, and it is
given by
\[ v([1,2],[1,2]) = \int_{\Hc_2}\prod_{j=1}^{2}\delta_D\left(
w_{j}([1,2])\right) \dd \yv \]
with $w_1([1,2]) = y_1-y_2$ and $w_2([1,2])=y_2-y_1$. It follows that
\[ v([1,2],[1,2]) =
\int_{\Hc_2}\delta_D(y_1-y_2)\delta_D(y_2-y_1) \dd \yv =1 \]
Finally,
\begin{eqnarray*}
v([1,2],[1,1])
&=& \int_{\Hc_2}\prod_{j=1}^{2} C\left(-\jj 2\pi\beta^{1/d}w_j([1,2])\right) \dd \yv \non
&=& \int_{\Hc_2} \left| C\left(-\jj
2\pi\beta^{1/d}(y_1-y_2)\right)\right|^2 \dd \yv
\end{eqnarray*}
Thus, we write
\[ \EE[\lambda_{d,\beta}^2] = 1 + \beta - \beta \left[\int_{\Hc_2} \left| C\left(-\jj
2\pi\beta^{1/d}(y_1-y_2)\right)\right|^2 \dd \yv\right]^d \]
}

\section{Convergence to the Mar\v{c}enko-Pastur distribution}
\label{sec:convergence}

In this section we show that the asymptotic eigenvalue distribution of
the matrix $\Tm_d$ tends to the Mar\v{c}enko-Pastur
law~\cite{MarcenkoPastur}, as $d \rightarrow \infty$. This is
equivalent to prove that, as $d \rightarrow \infty$, the $p$-th moment
of $\lambda_{d,\beta}$ tends to the $p$-th moment of the
Mar\v{c}enko-Pastur distribution with parameter $\beta$, for every $p
\geq 1$.

\begin{theorem}
\label{theorem2}
Let $\Tm_d$ be a $(2M+1)^d\times (2M+1)^d$ Hermitian random matrix as
defined in~(\ref{eq:t_d}), where the properties of the random vectors
$\xv_{\mu(\qv)}$ are described in Section~\ref{sec:qes_multi}. Let
$\EE[\lambda_{d,\beta}^p]$ be the $p$-th moment of the asymptotic
eigenvalue distribution of $\Tm_d$, given by Theorem~\ref{theorem1}.
Then, for any given $\beta$,

\begin{equation}
\lim_{d \rightarrow \infty} \EE[\lambda_{d,\beta}^p]
= \EE[\lambda_{\infty,\beta}^p] = \sum_{k=1}^p\beta^{p-k} N(p,k)
\label{eq:theorem2}
\end{equation}
where $N(p,k)$ are the {\em Narayana
  numbers}~\cite{EncIntSeq,Dumitriu} and
$\EE[\lambda_{\infty,\beta}^p]$ are the {\em Narayana polynomials},
i.e., the moments of the Mar\v{c}enko-Pastur
distribution~\cite{MarcenkoPastur}.
\end{theorem}
\begin{IEEEproof}

We first look at the expression of the $p$-th asymptotic moment and
observe that, for $h=k$, the contribution of the term in the right
hand side of~(\ref{eq:theorem1}) reduces to
\begin{equation}
 \sum_{k=1}^p \beta^{p-k}\sum_{\omegav \in
  \Omega_{p,k}}\sum_{\omegav' \in
  \Omega_{k,k}}u(\omegav')v(\omegav,\omegav')^d
\label{eq:contribution_h=k}
\end{equation}
The cardinality of $\Omega_{k,k}$ is $S(k,k)=1$ and $\Omega_{k,k} =
\{[1,\ldots,k]\}$. Thus, we only consider
$\omegav'=[1,\ldots,k]$. Moreover, using~(\ref{eq:u}) we have
$u([1,\ldots,k])=1$ since each subset $\Pc_{j'}([1,\ldots,k])$ has
cardinality $1$, $j'=1,\ldots,k$.  Therefore, the term
in~(\ref{eq:contribution_h=k}) becomes
\[ \sum_{k=1}^p\sum_{\omegav \in
  \Omega_{p,k}} \beta^{p-k} v(\omegav,[1,\ldots,k])^d \]
Using~(\ref{eq:v(omega,omega')}) with $h=k$, we have:
\begin{equation}
 v(\omegav,[1,\ldots,k])
 =\int_{\Hc_p}\prod_{j=1}^{k} \delta_D\left( w_{j}(\omegav)
   \right) \dd \yv \eqdef v(\omegav)
\end{equation}

Hence, the contribution to the $p$-th moment reduces to
\begin{equation}
 \sum_{k=1}^p \beta^{p-k}\sum_{\omegav \in
  \Omega_{p,k}} v(\omegav)^d
\label{eq:moments_h=k}
\end{equation}
In~\cite{izs08,TSP2_08} it is shown that, as $d \rightarrow \infty$,
~(\ref{eq:moments_h=k}) tends to the Narayana polynomial of order $p$.
It follows that, in order to prove the theorem, it is enough to show
that for $h<k$ the contribution of the term in the right hand side
of~(\ref{eq:theorem1}), to the expression of the $p$-th asymptotic
moment, vanishes as $d \rightarrow \infty$.  In practice we have to
show that, for each $\omegav \in \Omega_{p,k}$ and $\omegav' \in
\Omega_{k,h}$, with $h<k$,
\[ \lim_{d \rightarrow \infty} v(\omegav,\omegav')^d = 0 \]
or, equivalently, that $|v(\omegav,\omegav')|<1$.

We first notice that for $1<h<k$
\begin{eqnarray}
|v(\omegav,\omegav')|
&=& \left|\int_{\Hc_p}\prod_{j=1}^{k} C\left(-\jj 2\pi\beta^{1/d}w_j(\omegav)\right)
   \prod_{j'=1}^h \delta_D\left(\sum_{i'\in \Pc_{j'}(\omegav')}
   w_{i'}(\omegav)\right) \dd \yv \right| \non
&\le& \int_{\Hc_p}\left|\prod_{j=1}^{k} C\left(-\jj 2\pi\beta^{1/d}w_j(\omegav)\right)
   \prod_{j'=1}^h \delta_D\left(\sum_{i'\in \Pc_{j'}(\omegav')}
   w_{i'}(\omegav)\right) \right| \dd \yv \non
& =& \int_{\Hc_p}\prod_{j=1}^{k} \left|C\left(-\jj 2\pi\beta^{1/d}w_j(\omegav)\right)\right|
   \prod_{j'=1}^h \delta_D\left(\sum_{i'\in \Pc_{j'}(\omegav')}
   w_{i'}(\omegav)\right) \dd \yv
\label{eq:|v|}
\end{eqnarray}
Moreover, we have:
\begin{eqnarray}
\left| C\left(-\jj 2 \pi \beta^{1/d}w_j(\omegav)\right)\right|
&=& \left| \int_{-\infty}^{+\infty} \exp\left(-\jj 2 \pi
\beta^{1/d}w_j(\omegav) z\right)f_{\tilde{x}}(z) \dd z\right| \non
&\stackrel{(a)}{\le} & \int_{-\infty}^{+\infty}  \left| \exp\left(-\jj 2 \pi
\beta^{1/d}w_j(\omegav) z\right)f_{\tilde{x}}(z)\right| \dd x \non
&=& \int_{-\infty}^{+\infty}f_{\tilde{x}}(z) \dd z = 1
\end{eqnarray}
The equality $(a)$ arises if the condition $w_j(\omegav)=0$ is always
verified, otherwise, if $w_j(\omegav) \neq 0$, $\left| C\left(-\jj 2
\pi \beta^{1/d}w_j(\omegav)\right)\right| < 1$. 

Next, we make the
following observations: ({\em i}) since we consider partitions
$\omegav'$ of the form $\{1,\ldots,k\}$ in $h$ subsets with $h<k$,
then at least one of the sets $\Pc_{j'}(\omegav')$ has cardinality
$|\Pc_{j'}(\omegav')|>1$; ({\em ii}) the term
\[ \prod_{j'=1}^h\delta_D\left(\sum_{i'\in \Pc_{j'}(\omegav')}
w_{i'}(\omegav) \right) \] gives a non-zero contribution to the
integral in~(\ref{eq:|v|}) only when $\sum_{i'\in
  \Pc_{j'}(\omegav')}w_{i'}(\omegav)=0$.  Hence, if
$|\Pc_{j'}(\omegav')|>1$ for some $j'$, then some $w_{i'}(\omegav')
\neq 0$ will provide a non-zero contribution to the integral
in~(\ref{eq:|v|}).  In this case, we can write
\begin{eqnarray} |v(\omegav,\omegav')|
&\le& \int_{\Hc_p}\prod_{j=1}^{k} \left|C\left(-\jj 2\pi\beta^{1/d}w_{j}(\omegav)\right)\right|
   \prod_{j'=1}^h \delta_D\left(\sum_{i'\in \Pc_{j'}(\omegav')}
   w_{i'}(\omegav)\right) \dd \yv \non
&<& \int_{\Hc_p} \prod_{j'=1}^h\delta\left(\sum_{i'\in \Pc_{j'}(\omegav')} w_{i'}(\omegav) \right)\dd \yv \le 1
\end{eqnarray}
which proves the claim.

When $h=1$,  again, there is a measurable subset of $\Hc_p$ for which
$w_j(\omegav)\neq 0$, hence,
\[ |v(\omegav,\omegav')| \le \int_{\Hc_p}\prod_{j=1}^{k}
\left|C\left(-\jj 2\pi\beta^{1/d}w_j(\omegav)\right)\right| \dd \yv < 1 \]
i.e., the strict inequality holds.

\end{IEEEproof}

\insertfig{0.8}{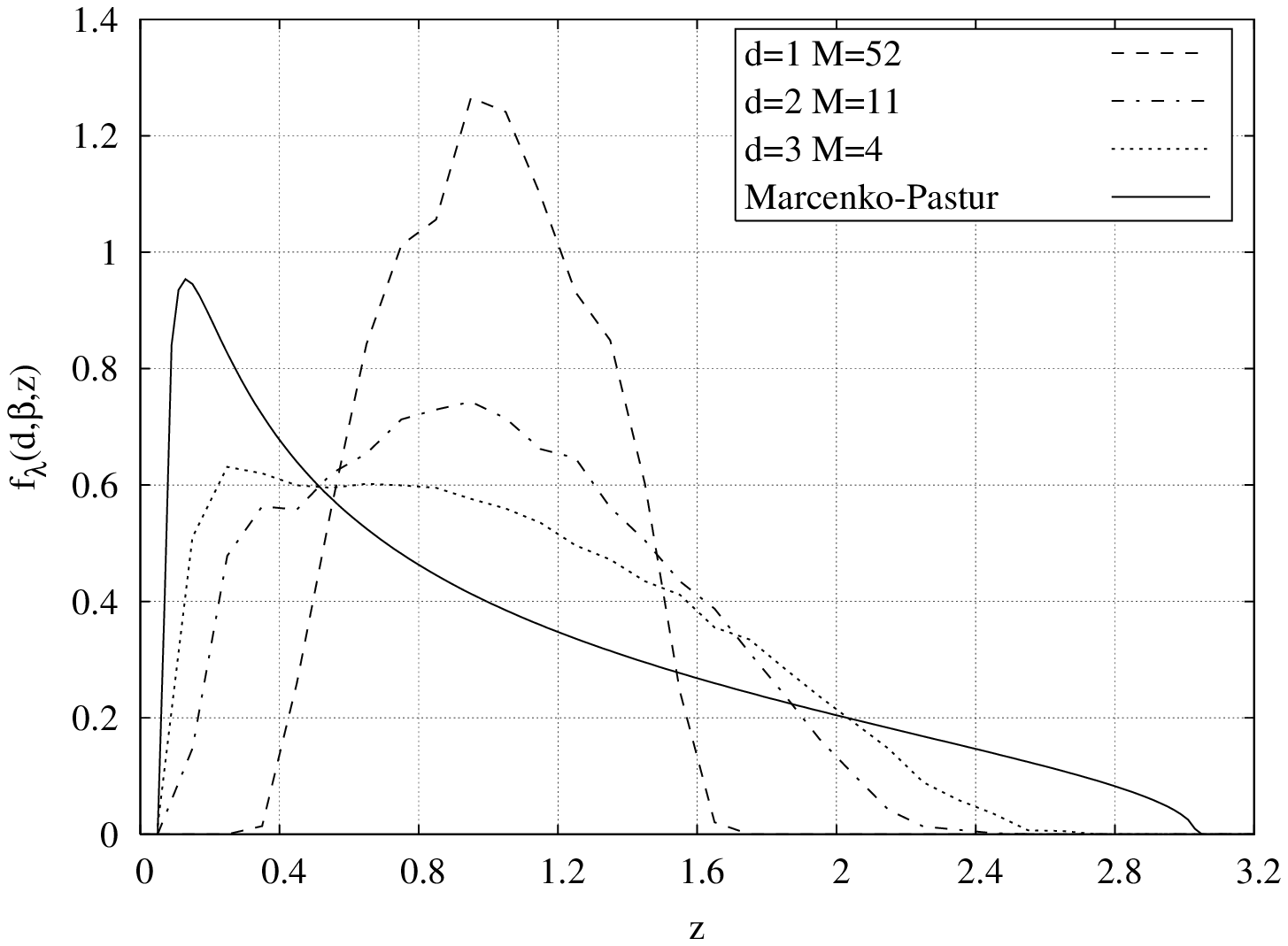}{Comparison between the
  Mar\v{c}enko-Pastur distribution and the empirical distribution
  obtained for $\beta=0.55$ and $d=1,2,3$ in the quasi equally
space case, and uniform $f_{\tilde{x}}(z)$}{fig:1} In
Figure~\ref{fig:1}, we show the empirical eigenvalue distribution of
the matrix $\Tm_d$ for $\beta=0.55$, $d=1,2,3$, and $\tilde{x}$
uniformly distributed in  $[0,1]$. The empirical
  distribution is compared to the Mar\v{c}enko-Pastur distribution
  (solid line).  We observe that as, $d$ increases, the
  Mar\v{c}enko-Pastur distribution law becomes a good approximation of
  $f_\lambda(d,\beta,z)$.  In particular, the two curves are
  relatively close for small $z$, already for $d=3$.

\section{Applications}
Here we present some applications where the results derived in this
work can be used.

The closed form expression of the moments of
  $f_\lambda(d,\beta,z)$, given by (\ref{eq:lambdap_8}), can be a
  useful basis for performing deconvolution operations, as proposed
  in~\cite{RyanDebbah2}.  As for the asymptotic approximation, we show
  below how to exploit our results for the estimation of the MSE
  provided by linear reconstruction techniques of irregularly sampled
  signals.

 Let us assume a general linear system model affected by additive
 noise.  For simplicity, consider a one-dimensional signal,
 $s(x)$. When observed over a finite interval, it admits an infinite
 Fourier series expansion~\cite{infocom07,TSP1_08}.  We can think of the
largest index $M$ of the non-negligible Fourier coefficients of the
expansion as the approximate one-sided bandwidth of the signal. We
therefore represent $s(x)$ by using $2M+1$ complex harmonics as
\begin{equation}
s(x) = \frac{1}{\sqrt{2M+1}}\sum_{k=-M}^M a_\ell\ee^{\jj 2 \pi \ell x}
\label{eq:s1}
\end{equation}
Now, consider that the signal is observed within one period interval
$[0,1)$ and sampled in $r$ points placed at positions
  $\xv=[x_0,\ldots,x_{r-1}]\Tran$, $x_q\in [0,1)$,
    $q=0,\ldots,r-1$. The complex numbers $a_\ell$ represent
    amplitudes and phases of the harmonics in $s(x)$. The signal
    samples $\sv=[s(x_0), \ldots, s(x_{r-1})]\Tran$ can be written as $\sv = \Gm\Herm\av $, where the matrix
    $\Gm$ is given in (\ref{eq:Gkq}).
    The signal discrete spectrum is given by the $2M+1$ complex vector
    $\av=[a_{-M}, \ldots, a_0, \ldots, a_M]^T$. We can now write the
    linear model for a measurement sample vector $\pv=[p(x_0), \ldots,
      p(x_{r-1})]\Tran$ taken at the sampling points $x_q$
\begin{equation}
\pv =\sv + \nv = \Gm\Herm\av + \nv \label{eq:lin_model}
\end{equation}
where $\nv$ is a random vector representing measurement noise. The
general problem is to reconstruct $\sv$ or $\av$ given the noisy
measurements $\pv$ \cite{izs08,TSP2_08}. A commonly used parameter to measure
the quality of the estimate of the reconstructed signal is the mean
square error (MSE).  In \cite{infocom07,TSP1_08,ipsn07} it has been shown
that, when linear reconstruction techniques are used and the sample
coordinates are known, the asymptotic MSE (i.e., as the number of
harmonics and the number of samples tend to infinity while their ratio
is kept constant) is a function of the asymptotic eigenvalue
distribution of the matrix $\Tm=\beta\Gm\Gm\Herm$, i.e.,
\begin{eqnarray}
\mbox{MSE} = \EE_{\lambda}\left[\frac{\beta}
{\lambda\, {\rm SNR}_m+\beta}\right]
\label{eq:mseinf_LMMSE3}
\end{eqnarray}
where the random variable $\lambda$ has distribution
$f_\lambda(d,\beta,z)$ and SNR$_m$ is the signal-to-noise ratio on
the measure. We therefore exploit our asymptotic approximation to
$f_\lambda(d,\beta,z)$  to compute (\ref{eq:mseinf_LMMSE3}).

\insertfig{0.8}{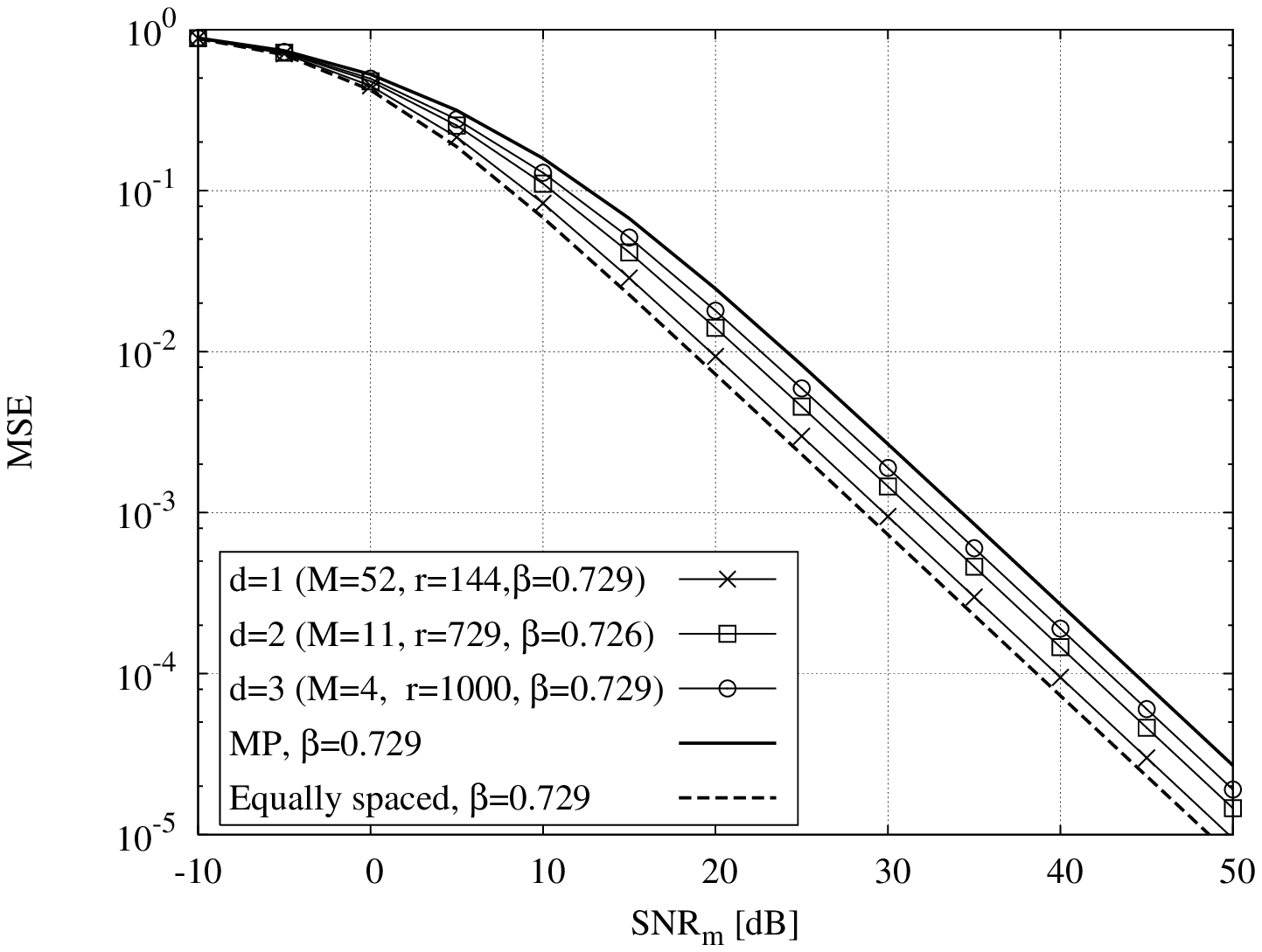}{MSE as a function of the
  signal-to-noise ratio  for $d=1,2,3$. The
  curves are compared with the results obtained through our asymptotic
analysis (MP) and with the equally spaced case}{fig:2}

\insertfig{0.8}{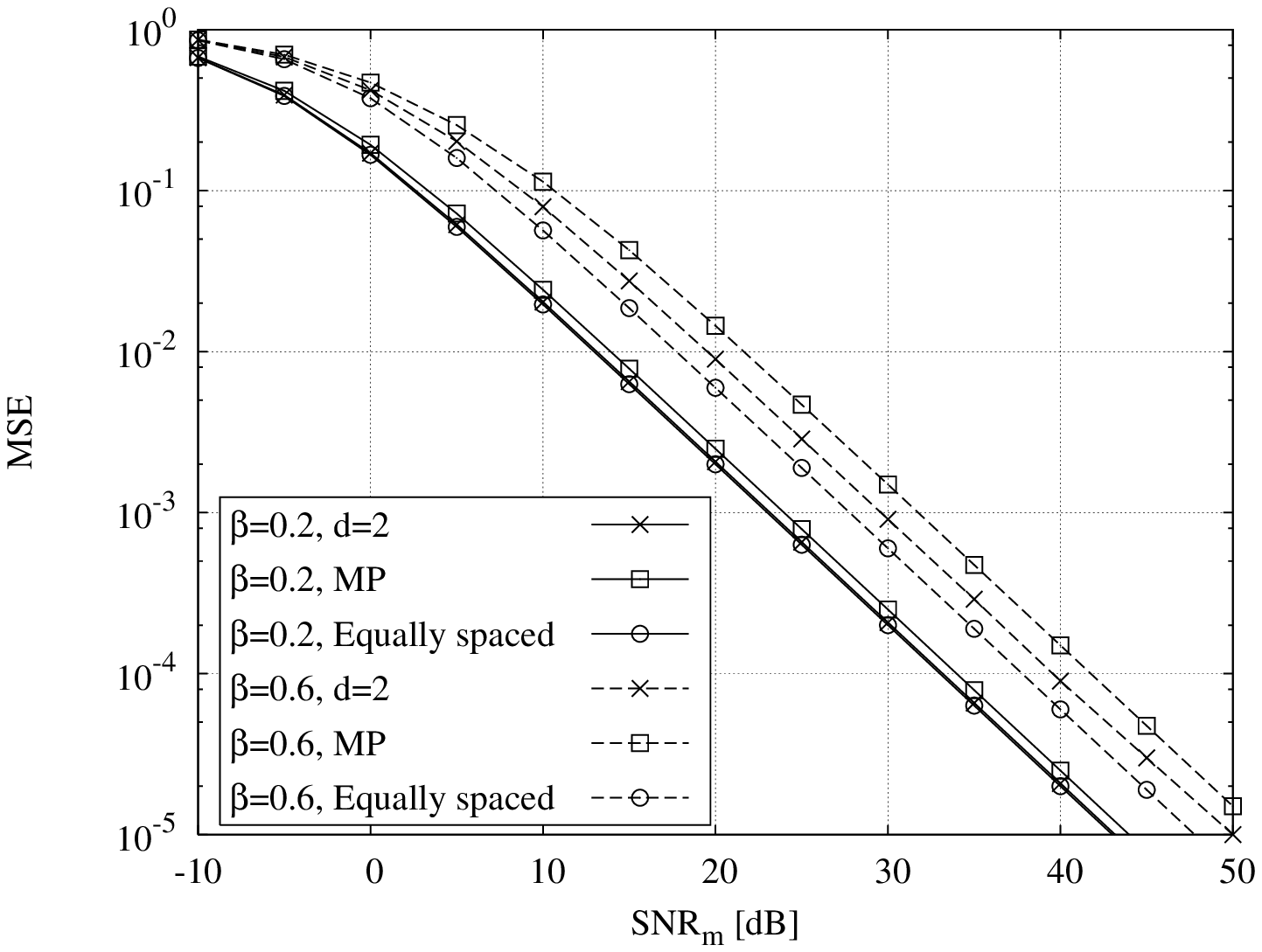}{MSE as a function of the
  signal-to-noise ratio  for $\beta=0.2, 0.6$. The
  curves are obtained for $d=2$ and compared against both
the equally spaced case and the results
derived through our asymptotic
analysis (MP)}{fig:3}

Figure~\ref{fig:2} shows the MSE obtained as a function of the
signal-to-noise ratio SNR$_m$.  The curves with markers labeled by
``$d=1,2,3$'' refer to the cases where the signal has dimension $d$
and the sampling points are quasi-equally spaced with jitter
$\tilde{x}$, uniformly distributed over $[0,1)$, and $\beta=0.729$.
The curve
  labeled by ``MP''  (thick line) reports the results
  derived through our asymptotic ($d \rightarrow \infty$)
  approximation to the eigenvalue distribution, while the curve
  labeled by ``Equally spaced'' (dashed line) represents the MSE
  achieved under a perfect equally spaced sample placement, i.e., when
  the eigenvalue distribution is given by $f_\lambda(d,\beta,z) =
  \delta_D(z-1)$. Notice that the MSE grows as $d$ increases and tends
  to the MSE obtained by a Mar\v{c}enko-Pastur eigenvalue
  distribution. Instead, as expected, the ``Equally spaced'' curve
  represents a lower bound to the system performance.

Figure~\ref{fig:3} presents similar results but obtained for $d=2$
and different values of $\beta$. We observe that the MSE obtained
through our asymptotic approximation (the curve labeled by ``MP'')
gives  excellent results for values of $\beta$ as small as 0.2, even
when compared against the numerical results derived by fixing $d=2$.
For $\beta=0.6$ (i.e., when the ratio of the number of signal
harmonics to the number of samples increases), the approximation
becomes slightly looser,  and the MSE computed by using
the Mar\v{c}enko-Pastur distribution gives an upper limit to the
quality of the reconstructed signal. Note that the smaller the
$\beta$, the higher  the oversampling rate relative to the equally
spaced minimal sampling rate $\beta=1$. We thus observe how our
bound becomes tighter as the oversampling rate increases.

To conclude, we describe some areas in signal processing 
where the above system model
and results find application.

\begin{itemize}
\item[{\em i)}] {\em Spectral estimation with noise.}  Spectral
  estimation from high precision sampling and quantization of
  bandlimited signals uses measurement systems which are usually
  affected by jitter \cite{Jenq97}. In such applications the
  quantization noise corresponds to the measurement noise and the
  jitter is caused by the limited accuracy of the timing circuits. In
  this case the sampling points are mismatched with respect to the
  nominal values, thus for $d=1$ we have: $x_q = \frac{q}{r} +
  \frac{\tilde{x}_q}{r}$ with some sampling rate $1/r$.  Note that
the exact positions of the samples are not known and  the
  case studied in this paper (i.e., MSE with exact positions) gives a
  lower bound to the reconstruction error.

\item[{\em ii)}] {\em Signal reconstruction in sensor networks.}
  Sensor networks, whose nodes sample a physical field, like air
  temperature, light intensity, pollution levels or rain falls,
  typically represent an example of quasi-equally spaced sampling
  \cite{ipsn07,Ganesan,Zhao,Early}. Indeed, often sensors are not
  regularly deployed in the area of interest due to terrain conditions
  and deployment practicality and, thus, the physical field is not
  regularly sampled in the space domain. Sensors report the data to a
  common processing unit (or {\em sink} node), which is in charge of
  reconstructing the sensed field, based on the received samples and
  on the knowledge of their coordinates. If the field can be
  approximated as bandlimited in the space domain, then an estimate of
  the discrete spectrum can be obtained by using linear reconstruction
  techniques \cite{Feichtinger95,ipsn07}, even in presence of additive
  noise.  In this case, our approximation allows to compute the MSE on the
  reconstructed field.

\item[{\em iii)}] {\em Stochastic sampling in computer graphics and
  image processing.}  Jittered sampling was first
  examined by Balakrishnan in \cite{Balakrishnan62}, who analyzed it
  as an undesirable effect in sampling continuous time functions. More
  than twenty years later, Cook \cite{Cook86} realized that the effect
  of stochastic sampling can be advantageous in computer graphics to
  reduce aliasing artifacts, and considered jittering a regular grid
  as an effective sampling technique.  Another example of sampling
  with jitter was recently proposed in
  \cite{Marziliano06}, for robust authentication of images.

\end{itemize}

\section{Conclusions}
We studied the behavior of the eigenvalue distribution of a class of
random matrices, which find large application in signal and image
processing.  In particular, by using asymptotic analysis, we derived
a closed-form expression for the moments of the eigenvalue
distribution.
Using these moments, we showed that, as the signal dimension goes to
infinity, the asymptotic eigenvalue distribution tends to the
Mar\v{c}enko-Pastur law.  This result allowed us to obtain a simple
and accurate bound to the signal reconstruction error,
which can find application in several fields, such as jittered
sampling, sensor networks, computer graphics and image processing.

\appendices

\section{Proof of Theorem~\ref{theorem1}}
\label{app:proof_theorem1}
Using~(\ref{eq:t_d}), the term $\trace\EE_{\Xc}\left[\Tm_d^p\right]$
in~(\ref{eq:lambda_p}) can be written as:
\begin{eqnarray}
\trace\EE_{\Xc}\left[\Tm_d^p\right]
& = & \EE_{\Xc}\left[\sum_{\ellv_1}(\Tm_d^p)_{\nu(\ellv_1),\nu(\ellv_1)}\right] \non
& = & \EE_{\Xc}\left[\sum_{\ellv_1}\cdots\sum_{\ellv_p}(\Tm_d)_{\nu(\ellv_1),\nu(\ellv_2)}\cdots(\Tm_d)_{\nu(\ellv_p),\nu(\ellv_1)}\right]\non
& = & \frac{1}{r^p}\sum_{\ellv_1}\cdots\sum_{\ellv_p}\sum_{\qv_1}\cdots\sum_{\qv_p} \EE_{\Xc}\left[\exp\left(-\jj 2\pi  \sum_{i=1}^p
\xv_{\mu(\qv_i)}\Tran(\ellv_i-\ellv_{[i+1]})\right)\right] \non
& = & \frac{1}{r^p}\sum_{\Lm \in \Lc_d}\sum_{\Qm \in \Qc_d}\EE_{\Xc}\left[\exp\left(-\jj 2\pi  \sum_{i=1}^p
\xv_{\mu(\qv_i)}\Tran(\ellv_i-\ellv_{[i+1]})\right)\right]
\label{eq:lambda_p2}
\end{eqnarray}
where $\Qc_d$ and $\Lc_d$ are sets of integer matrices such that
\begin{eqnarray*}
\Qc_d &=& \left\{\Qm~|~\Qm=[\qv_1, \ldots,
  \qv_p],~~\qv_i=[q_{i,1},\ldots, q_{i,d}]\Tran, q_{i,m}=0,\ldots,\rho-1\right\} \\
\Lc_d &=& \left\{\Lm \,|\, \Lm=[\ellv_1, \ldots, \ellv_p],~~\ellv_i=[\ell_{i,1},\ldots,
\ell_{i,d}]\Tran, \ell_{i,m}=-M,\ldots M\right\}
\end{eqnarray*}
and
\[ [i+1] = \left\{\begin{array}{ll} i+1 & 1\le i < p \\ 1  & i=p \end{array}\right. \]
\subsection{Set partitioning}
\label{sec:set_partitioning}
We now apply the definitions in Section~\ref{sec:definitions} in order
to rewrite~(\ref{eq:lambda_p2}) using set partitioning. In particular
by considering the vector $\muv=\muv(\Qm) \eqdef [\mu_1, \ldots,\mu_p]\Tran$
where $\mu_i =\mu(\qv_i)$ and $\qv_i$ is the $i$-th column of
$\Qm$, we observe that:
\begin{itemize}
\item the vector $\muv$ is uniquely defined by $\Qm$, and a given
  $\muv$ uniquely defines a matrix $\Qm \in {\Qc}_d$ since $\mu(\cdot)$
  is an invertible function;
\item a given $\muv$ induces a partition $\omegav(\muv)$;
\item since $r$ is the number of values that the entries $\mu_i$ can
   take, there exist $r!/(r-k(\muv))!$ matrices $\Qm\in \Qc_d$
   generating a given partition of $\Pc$ made of $k(\muv)$
    subsets. In other words $r!/(r-k(\muv))!$ distinct $\muv$'s yield
    the same partition $\omegav(\muv)$.
\end{itemize}

Since the random vectors $\xv_{\mu(\qv')}$ and $\xv_{\mu(\qv'')}$ are
independent for $\qv'\neq \qv''$, for any given $\Qm$ the average operator
in (\ref{eq:lambda_p2}) factorizes into $k(\muv)$ terms, i.e.,
\begin{eqnarray}
\EE_{\Xc}\left[\exp\left(- \jj 2\pi \sum_{i=1}^p \xv_{\mu(\qv_i)}\Tran(\ellv_i-\ellv_{[i+1]})\right)\right]
&=& \EE_{\Xc}\left[\exp\left(- \jj 2\pi \sum_{i=1}^p \xv_{\mu_i}\Tran (\ellv_i-\ellv_{[i+1]})\right)\right] \non
&=& \prod_{j=1}^{k(\muv)} \EE_{\xv_{\gamma_j}}\left[\exp\left( -\jj
  2\pi  \xv_{\gamma_j}\Tran \sum_{i \in
    \Pc_j(\muv)}\ellv_i-\ellv_{[i+1]} \right)\right] \non
&=& \prod_{j=1}^{k(\muv)}
\EE_{\xv_{\gamma_j}}\left[\zeta^{\rho \xv_{\gamma_j}\Tran\hat{\wv}_j(\muv)}\right]
\label{eq:factorize1}
\end{eqnarray}
indeed, for every $i \in \Pc_j(\muv)$, we have $\mu_i = \gamma_j$.
In the last line of~(\ref{eq:factorize1}), we exploited the following
two definitions
\[ \zeta = \exp(-\jj 2 \pi/\rho) \]
and
\begin{equation}
\hat{\wv}_j(\muv) = \sum_{i \in \Pc_j(\muv)}\ellv_i-\ellv_{[i+1]}
\label{eq:w}
\end{equation}
Also, note that, in the product in~(\ref{eq:factorize1}), each factor depends on a single
random vector, $\xv_{\gamma_j}$. Since
$\xv_{\mu(\qv)}=\qv/\rho+\tilde{\xv}_{\mu(\qv)}/\rho$ and
$\mu(\cdot)$ is invertible then, by defining $\bar{\xv}_{\gamma_j} =
\mu^{-1}(\gamma_j)$ we have
\[ \xv_{\gamma_j}=\bar{\xv}_{\gamma_j}/\rho+\tilde{\xv}_{\gamma_j}/\rho \]
and
\begin{equation}
\EE_{\xv_{\gamma_j}}\left[ \zeta^{\rho \xv_{\gamma_j}\Tran\hat{\wv}_j(\muv)}\right]
= \zeta^{\bar{\xv}_{\gamma_j}\Tran\hat{\wv}_j(\muv)}
\EE_{\tilde{\xv}_{\gamma_j}}\left[ \zeta^{\tilde{\xv}_{\gamma_j}\Tran\hat{\wv}_j(\muv)} \right]
= \zeta^{\bar{\xv}_{\gamma_j}\Tran\hat{\wv}_j(\muv)}
\EE_{\tilde{\xv}}\left[\zeta^{\tilde{\xv}\Tran\hat{\wv}_j(\muv)} \right]
\label{eq:split}
\end{equation}
In the last term of~(\ref{eq:split}) we removed the subscript $\gamma_j$
from the argument of the average operator, since the distribution of
$\tilde{\xv}_{\gamma_j}$ does not depend on $\gamma_j$.
Summarizing, the term $\trace\EE_{\Xc}\left[\Tm_d^p\right]$ in~(\ref{eq:lambda_p}) can be
written as
\begin{eqnarray}
\trace\EE_{\Xc}\left[\Tm_d^p\right]
&=& \frac{1}{r^p}~\sum_{\Qm \in \Qc_d}~\sum_{\Lm \in
  \Lc_d}\prod_{j=1}^{k(\muv)}
\zeta^{\bar{\xv}_{\gamma_j}\Tran\hat{\wv}_j(\muv)}
\EE_{\tilde{\xv}}\left[ \zeta^{\tilde{\xv}\Tran\hat{\wv}_j(\muv)}\right] \non
\label{eq:lambda_p4}
\end{eqnarray}
Since each $\Qm$ is uniquely identified by a vector $\muv$,
we can observe that
\begin{equation}
 \sum_{\Qm \in \Qc_d} f(\muv)
  = \sum_{\omegav \in \Omega_p}\sum_{\muv \in \Mc(\omegav)}
 f(\muv) = \sum_{k=1}^p \sum_{\omegav \in \Omega_{p,k}}\sum_{\muv \in
   \Mc(\omegav)} f(\muv)
\label{eq:sums}
\end{equation}
for every function $f(\muv)$. Recall that, in~(\ref{eq:sums}),
$\Mc(\omegav)$ represents the set of $\muv$ inducing a given partition
$\omegav$.

From the definitions in Section~\ref{sec:definitions}, it follows that,
if $\muv$ induces $\omegav$, then $k(\muv)=k(\omegav)$,
$\Pc_j(\muv) = \Pc_j(\omegav)$, and $\hat{\wv}_j(\muv) = \hat{\wv}_j(\omegav)$,
$j=1,\ldots,k(\omegav)$. Therefore,
\begin{eqnarray}
\trace\EE_{\Xc}\left[\Tm_d^p\right]
&=& \frac{1}{r^p}~\sum_{k=1}^p \sum_{\omegav \in
  \Omega_{p,k}}\sum_{\muv \in \Mc(\omegav)}
\sum_{\Lm \in \Lc_d}\prod_{j=1}^{k} \zeta^{\bar{\xv}_{\gamma_j}\Tran\hat{\wv}_j(\muv)}\EE_{\tilde{\xv}}\left[\zeta^{\tilde{\xv}\Tran\hat{\wv}_j(\muv)} \right] \non
&=& \frac{1}{r^p}~\sum_{k=1}^p \sum_{\omegav \in
  \Omega_{p,k}}\sum_{\muv \in \Mc(\omegav)}
\sum_{\Lm \in \Lc_d}\prod_{j=1}^{k}\zeta^{\bar{\xv}_{\gamma_j}\Tran\hat{\wv}_j(\omegav)}
    \EE_{\tilde{\xv}}\left[ \zeta^{\tilde{\xv}\Tran\hat{\wv}_j(\omegav)}\right] \non
&=& \frac{1}{r^p}~\sum_{k=1}^p \sum_{\omegav \in \Omega_{p,k}}
\sum_{\Lm \in \Lc_d}\sum_{\muv \in
  \Mc(\omegav)}\left[\prod_{j=1}^{k}\zeta^{\bar{\xv}_{\gamma_j}\Tran\hat{\wv}_j(\omegav)}\right]
\left[\prod_{j=1}^{k}\EE_{\tilde{\xv}}\left[\zeta^{\tilde{\xv}\Tran\hat{\wv}_j(\omegav)} \right]\right] \non
&\stackrel{(a)}{=}& \frac{1}{r^p}~\sum_{k=1}^p\sum_{\omegav \in \Omega_{p,k}}
\sum_{\Lm \in \Lc_d}  \eta(\omegav,\Lm)
\sum_{\muv \in \Mc(\omegav)} \prod_{j=1}^{k}
  \zeta^{\bar{\xv}_{\gamma_j}\Tran\hat{\wv}_j(\omegav)}
\label{eq:lambda_p5}
\end{eqnarray}
In~(\ref{eq:lambda_p5}) we defined
\begin{equation}
\eta(\omegav,\Lm) =
\prod_{j=1}^{k}\EE_{\tilde{\xv}}\left[\zeta^{\tilde{\xv}\Tran\hat{\wv}_j(\omegav)}\right] =
\prod_{j=1}^{k}\prod_{m=1}^d \EE_{\tilde{x}_m}\left[\zeta^{\tilde{x}_m \hat{w}_{jm}(\omegav)}\right]
\label{eq:eta}
\end{equation}
where $\tilde{x}_m$ and $\hat{w}_{jm}$ are the $m$-th entries of
$\tilde{\xv}$ and $\hat{\wv}_{j}$, respectively.
In the equality ``(a)'' we exploited the fact that the term
$\zeta^{\tilde{\xv}\Tran\hat{\wv}_j(\omegav)}$ does not depend on $\muv$ and
can be factored from the sum over $\muv$.
As for the term $\sum_{\muv \in \Mc(\omegav)} \prod_{j=1}^{k}
   \zeta^{\bar{\xv}_{\gamma_j}\Tran\hat{\wv}_j}$, we have the following lemma.

\begin{lemma}
\label{lemma:1}
Let $\omegav \in \Omega_{p,k}$, let $\hat{\wv}_1, \ldots, \hat{\wv}_k$ be
vectors of size $d$ with integer entries, defined as in~(\ref{eq:w}). Let
$\Mc(\omegav)$ be the set of vectors $\muv$ inducing $\omegav$. Then
\begin{eqnarray}
\sum_{\muv \in \Mc(\omegav)} \prod_{j=1}^{k}
   \zeta^{\bar{\xv}_{\gamma_j}\Tran\hat{\wv}_j}
&=& \sum_{h=1}^k r^h  \sum_{\omegav' \in \Omega_{k,h}}
    u(\omegav')\prod_{j'=1}^h\delta\left(\sum_{i'\in
     \Pc_{j'}(\omegav')} \hat{\wv}_{i'}(\omegav) \right)
\end{eqnarray}
where $u(\omegav')=(-1)^{k-h}\prod_{j'=1}^h (|\Pc_{j'}(\omegav')|-1)!$, $\gamma_j = \gamma_j(\muv)$, and where $\Omega_{k,h}$ is the set
of vectors $\omegav'$ of size $k$, representing the partitions of the
set $\Pc' = \{1,\ldots,k\}$ in $h$ subsets, namely,
$\Pc_1'(\omegav'),\ldots,\Pc_h'(\omegav')$.
\end{lemma}

\begin{IEEEproof}
The proof can be found in Appendix~\ref{app:proof_lemma1}.
\end{IEEEproof}

By applying the result of Lemma~\ref{lemma:1} to~(\ref{eq:lambda_p5}),
we get
\begin{eqnarray}
\trace\EE_{\Xc}\left[\Tm_d^p\right]
&=& \sum_{k=1}^p\sum_{h=1}^k\sum_{\omegav \in \Omega_{p,k}} \sum_{\omegav' \in \Omega_{k,h}}
\frac{r^h u(\omegav')}{r^p}\sum_{\Lm \in \Lc_d}  \eta(\omegav,\Lm)
 \prod_{j'=1}^h \delta\left(\sum_{i'\in \Pc_{j'}(\omegav')} \hat{\wv}_{i'}(\omegav) \right)
\label{eq:lambda_p6}
\end{eqnarray}
Considering that
\[ \prod_{j'=1}^h\delta\left(\sum_{i'\in \Pc_{j'}(\omegav')} \hat{\wv}_{i'}(\omegav) \right)=\prod_{j'=1}^h\prod_{m=1}^d\delta\left(\sum_{i'\in \Pc_{j'}(\omegav')} \hat{w}_{i'm}(\omegav) \right)\]
and by using~(\ref{eq:eta}) and~(\ref{eq:lambda_p6}), we have
\begin{eqnarray}
&&\sum_{\Lm \in \Lc_d} \eta(\omegav,\Lm) \prod_{j'=1}^h \delta\left(\sum_{i'\in
   \Pc_{j'}(\omegav')} \hat{\wv}_{i'}(\omegav) \right) \non
&&\qquad =
\sum_{\ellv_1 \in \Lc_1}\cdots\sum_{\ellv_d \in
  \Lc_1}\prod_{j=1}^{k}\prod_{m=1}^{d}\EE_{\tilde{x}_m}\left[
  \zeta^{\tilde{x}_m \hat{w}_{jm}(\omegav)}\right]
 \prod_{j'=1}^h \prod_{m=1}^{d}\delta\left(\sum_{i'\in
   \Pc_{j'}(\omegav')} \hat{w}_{i'm}(\omegav) \right) \non
&&\qquad = \left[\sum_{\ellv \in \Lc_1}
   \prod_{j=1}^{k}\EE_{\tilde{x}}\left[ \zeta^{\tilde{x} \hat{w}_j(\omegav)}\right]
 \prod_{j'=1}^h \delta\left(\sum_{i'\in
   \Pc_{j'}(\omegav')} \hat{w}_{i'}(\omegav) \right)\right]^d = \psi_M(\omegav,\omegav')^d
\end{eqnarray}
where the subscript $_M$ highlights the dependency of $\ellv$ on $M$.

In conclusion,
\begin{eqnarray}
\trace\EE_{\Xc}\left[\Tm_d^p\right]
&=&\sum_{k=1}^p\sum_{h=1}^k \sum_{\omegav \in \Omega_{p,k}}\sum_{\omegav' \in \Omega_{k,h}}
\frac{r^h u(\omegav')}{r^p} \psi_M(\omegav,\omegav')^d
\label{eq:trace_8}
\end{eqnarray}
To compute $\EE[\lambda_{d,\beta}^p]$, we consider the
limit in~(\ref{eq:lambda_p}). By using the
definition~(\ref{eq:beta_qes}), we first notice that
\[ \frac{r^h}{r^p(2M+1)^d} =
\frac{\beta^{p-h}}{(2M+1)^{d(p-h+1)}} \]
Then, by using~(\ref{eq:trace_8})  in~(\ref{eq:lambda_p}), we obtain
\begin{eqnarray}
\EE[\lambda_{d,\beta}^p]
&=& \limbeta \sum_{k=1}^p\sum_{h=1}^{k}\frac{\beta^{p-h}}{(2M+1)^{d(p-h+1)}}\sum_{\omegav \in \Omega_{p,k}} \sum_{\omegav' \in \Omega_{k,h}}
u(\omegav')\psi_M(\omegav,\omegav')^d \non
&=&\sum_{k=1}^p\sum_{h=1}^{k} \beta^{p-h}\sum_{\omegav \in \Omega_{p,k}}\sum_{\omegav' \in \Omega_{k,h}}u(\omegav')
 \left[\lim_{M\rightarrow \infty}
   \frac{\psi_M(\omegav,\omegav')}{(2M+1)^{p-h+1}}\right]^d \non
&=&\sum_{k=1}^p\sum_{h=1}^{k} \beta^{p-h}\sum_{\omegav \in \Omega_{p,k}}\sum_{\omegav' \in \Omega_{k,h}}u(\omegav')
 v(\omegav,\omegav')^d
\label{eq:lambdap_8}
\end{eqnarray}
The second equality in~(\ref{eq:lambdap_8}) holds since, for any given
$p$, the sums $\sum_{\omegav \in \Omega_{p,k}}$ and $\sum_{\omegav' \in
  \Omega_{k,h}}$ are over a finite number of terms, and the coefficients
$u(\omegav')$ are finite and do not depend on $M$. Therefore, the
limit operator can be swapped with the summations.
The coefficient $v(\omegav,\omegav')$ is defined as
\begin{eqnarray}
v(\omegav,\omegav')
&=&\lim_{M\rightarrow \infty}
\frac{\psi_M(\omegav,\omegav')}{(2M+1)^{p-h+1}}\non
&=& \lim_{M\rightarrow \infty}
\frac{1}{(2M+1)^{p-h+1}}\sum_{\ellv \in \Lc_1}
   \prod_{j=1}^{k}\EE_{\tilde{x}}\left[\zeta^{\tilde{x} \hat{w}_j(\omegav)}\right]
 \prod_{j'=1}^h \delta\left(\sum_{i'\in \Pc_{j'}(\omegav')} \hat{w}_{i'}(\omegav)
 \right) \non
&\stackrel{(a)}{=}& \lim_{M\rightarrow \infty}
\frac{1}{(2M+1)^{p-h+1}}\sum_{\ellv \in \Lc_1}
   \prod_{j=1}^{k} C\left(-\jj 2 \pi \hat{w}_j(\omegav)/\rho\right)
   \prod_{j'=1}^h \delta\left(\sum_{i'\in \Pc_{j'}(\omegav')}
   \hat{w}_{i'}(\omegav)\right)
\label{eq:v2}
\end{eqnarray}
where, in the equality $(a)$,
we introduced the characteristic function of
$\tilde{x}$, defined as $C(s) = \EE_{\tilde{x}}[\ee^{sz}]$.
We now consider three possible cases:
\begin{itemize}
\item if $h=1$, then $\Omega_{k,1}=\{[\underbrace{1,\ldots,1}_{k}]\}$,
  thus we only consider $\omegav'=[\underbrace{1,\ldots,1}_{k}]$.
  Then, $\Pc_1(\omegav')=\{1,\ldots,k\}$ and
\begin{eqnarray}
\sum_{i' \in \Pc_1(\omegav')} \hat{w}_{i'}(\omegav)
&=& \sum_{i' \in \{1,\ldots,k\}} \hat{w}_{i'}(\omegav) \non
&=& \sum_{i'=1}^k \hat{w}_{i'}(\omegav) \non
&=& \sum_{i'=1}^k \sum_{i \in \Pc_{i'}(\omegav)}\ell_i-\ell_{[i+1]} \non
&=& \sum_{i=1}^p \ell_i-\ell_{[i+1]} = 0
\end{eqnarray}
and by consequence $\delta\left(\sum_{i'\in
  \Pc_{j'}(\omegav')}\hat{w}_{i'}(\omegav)\right)=1$.
Hence,
\begin{equation}
 v(\omegav,\omegav') = \int_{\Hc_p}\prod_{j=1}^{k} C\left(-\jj 2\pi\beta^{1/d}w_j(\omegav)\right)
    \dd \yv
 \label{eq:v_h=1}
\end{equation}
where, in analogy with~(\ref{eq:w}), we defined
\[ w_j = \sum_{i \in \Pc_j(\omegav)} y_i-y_{[i+1]} \]
$y_i\in \RR$, $i=1,\ldots,p$.
We denote by $\yv$ the vector $\yv=[y_1,\ldots,y_p]\Tran$;
\item if $1<h<k$, the argument of the $\delta(\cdot)$ function in~(\ref{eq:v2})
is always a function of the indices $\ell_i$. Thus
\[ \int_{\Hc_p}\prod_{j=1}^{k} C\left(-\jj 2\pi\beta^{1/d}w_j(\omegav)\right)
   \prod_{j'=1}^h \delta_D\left(\sum_{i'\in \Pc_{j'}(\omegav')}
   w_{i'}(\omegav)\right) \dd \yv \]
where $\delta_D(\cdot)$ denotes the Dirac's delta;
\item if $h=k$, the cardinality of $\Omega_{k,h}=\Omega_{k,k}$ is $S(k,k)=1$
and $\Omega_{k,k} = \{[1,\ldots,k]\}$. Thus, we only consider
$\omegav'=[1,\ldots,k]$.
It follows that:
\begin{eqnarray}
 v(\omegav,\omegav')
 &=&\int_{\Hc_p}\prod_{j=1}^{k} C\left(-\jj 2\pi\beta^{1/d}w_j(\omegav)\right)
   \prod_{j'=1}^k \delta_D\left(\sum_{i'\in \Pc_{j'}([1,\ldots,k])} w_{i'}(\omegav)\right) \dd \yv \non
 &=&\int_{\Hc_p}\prod_{j=1}^{k} C\left(-\jj 2\pi\beta^{1/d}w_j(\omegav)\right)
   \delta_D\left(\sum_{i'\in \Pc_{j}([1,\ldots,k])} w_{i'}(\omegav)\right) \dd \yv
\end{eqnarray}
Since $\Pc_j([1,\ldots,k])= \{j\}$ and $C(0)=1$, we have
\begin{eqnarray}
 v(\omegav,[1,\ldots,k])
 &=&\int_{\Hc_p}\prod_{j=1}^{k} C\left(-\jj 2\pi\beta^{1/d}w_j(\omegav)\right)
   \delta_D\left( w_{j}(\omegav) \right) \dd \yv \non
 &=&\int_{\Hc_p}\prod_{j=1}^{k} C(0)
   \delta_D\left( w_{j}(\omegav) \right) \dd \yv \non
 &=&\int_{\Hc_p}\prod_{j=1}^{k} \delta_D\left( w_{j}(\omegav)
   \right) \dd \yv
\end{eqnarray}
\end{itemize}
As a last remark, if $k=1$, we have $h=1$ and $\Omega_{p,k}
=\Omega_{p,1}=\{[\underbrace{1,\ldots,1}_{p}]\}$.
Then $w_j(\omegav) = \sum_{i=1}^p w_i = 0$.
Using~(\ref{eq:v_h=1}), we obtain
\[ v(\omegav,\omegav') = \int_{\Hc_p}\prod_{j=1}^{k} C\left(-\jj 2\pi\beta^{1/d}w_j(\omegav)\right)
    \dd \yv = \int_{\Hc_p}\prod_{j=1}^{k} C(0)\dd \yv = 1 \]

\section{Proof of Lemma~\ref{lemma:1}}
\label{app:proof_lemma1}
Recall that $\Mc(\omegav)$ denotes the set of vectors
$\muv=[\mu_1,\ldots,\mu_p]$ inducing the same partition $\omegav$.  As
defined in Section~\ref{sec:definitions}, if $\omegav \in \Omega_{p,k}$,
then each $\muv \in \Mc(\omegav)$ contains $k$ distinct values, namely,
$\gammav = [\gamma_1,\ldots,\gamma_k]$ where $0\le \gamma_j <r$,
$j=1,\ldots,k$ and $\gamma_j \neq \gamma_{j'}$ for each
$j,j'=1,\ldots,k$ and $j\neq j'$.  Therefore, from~(\ref{lemma:1}) we
can write
\[
\sum_{\muv \in \Mc(\omegav)} \prod_{j=1}^{k}
\zeta^{\bar{\xv}_{\gamma_j}\Tran\hat{\wv}_j} =
\sum_{\substack{\gamma_1,\ldots,\gamma_k \\ \neq}} \prod_{j=1}^{k}
\zeta^{\bar{\xv}_{\gamma_j}\Tran\hat{\wv}_j} \] where the symbol
$\sum_{\substack{\gamma_1,\ldots,\gamma_k \\ \neq}}$ indicates a sum
over the variables $\gamma_1,\ldots,\gamma_k$ with the constraint that
$\gamma_j \neq \gamma_{j'}$ for every $j,j'=1,\ldots,k$ and $j\neq j'$.
Notice that the values $\gamma_j$ ($j=1,\ldots,k$) are the scalar
counterparts of the integer vectors $\vv_1,\ldots,\vv_k$, $\vv_j =
[v_{j1}, \ldots,v_{jd}]\Tran$, $0 \le v_{jm} < \rho$, $m=1,\ldots,d$,
through the invertible function $\mu(\cdot)$, i.e., $\gamma_j =
\mu(\vv_j)$, $j=1,\ldots,k$.  Hence, by definition of $\bar{\xv}$,
we have $\bar{\xv}_{\gamma_j} = \bar{\xv}_{\mu(\vv_j)} = \vv_j$ and in
conclusion
\begin{equation}
\label{eq:proof_lemma1_1} \sum_{\muv \in \Mc(\omegav)} \prod_{j=1}^{k}
\zeta^{\bar{\xv}_{\gamma_j}\Tran\hat{\wv}_j} =
\sum_{\substack{\vv_1,\ldots,\vv_k \\ \neq}}
\prod_{j=1}^{k}\zeta^{\vv_j\Tran\hat{\wv}_j} =
\sum_{\substack{\vv_1,\ldots,\vv_k \\ \neq}}
\zeta^{\vv_1\Tran\hat{\wv}_1+ \cdots + \vv_k\Tran\hat{\wv}_k}
\end{equation}

We now compute the last term of~(\ref{eq:proof_lemma1_1}) by summing over one variable
at a time. We first notice that, for every set $\vv_1,\ldots, \vv_n$ of
distinct vectors
\[ \sum_{\vv \neq \vv_1, \ldots, \vv_n} \zeta^{\vv\Tran\hat{\wv}}  = \left\{ \begin{array}{ll} r-n
  & \hat{\wv}=\zerov \\ -\sum_{j=1}^n \zeta^{\vv_j\Tran\hat{\wv}} &
  \hat{\wv} \neq \zerov \end{array}\right. \]
In particular when $\wv \neq \zerov$, $\sum_{\vv} \zeta^{\vv\Tran\hat{\wv}}=0$.

Let us arbitrarily choose
the variable $\vv_k$. If by hypothesis $\wv_k\neq \zerov$, then by
summing~(\ref{eq:proof_lemma1_1}) over $\vv_k$ we get
\begin{equation}
\label{eq:proof_lemma1_2}
\sum_{\substack{\vv_1,\ldots,\vv_k \\ \neq}}
\zeta^{\vv_1\Tran\hat{\wv}_1+ \cdots + \vv_k\Tran\hat{\wv}_k} =
-\sum_{j=1}^{k-1} \sum_{\substack{\vv_1,\ldots,\vv_{k-1} \\ \neq}}
\zeta^{\vv_1\Tran\hat{\wv}_1+ \cdots +
  \vv_{k-1}\Tran\hat{\wv}_{k-1}}\zeta^{\vv_j\Tran\hat{\wv}_k}
\end{equation}
We compute separately each of the $k-1$ contributions
in~(\ref{eq:proof_lemma1_2}). In particular, the generic $j'$-th term
($j=j'$) is given by
\[ -\sum_{\substack{\vv_1,\ldots,\vv_{k-1} \\ \neq}}
    \zeta^{\vv_1\Tran\hat{\wv}_1+ \cdots +
      \vv_{k-1}\Tran\hat{\wv}_{k-1}}\zeta^{\vv_{j'}\Tran\hat{\wv}_k} =
    -\sum_{\substack{\vv_1,\ldots,\vv_{k-1} \\ \neq}}
    \zeta^{\vv_1\Tran\hat{\wv}_1+ \cdots +
      \vv_{j'}\Tran(\hat{\wv}_{j'}+\hat{\wv}_k)+
      \vv_{k-1}\Tran\hat{\wv}_{k-1}} \]

We now proceed by summing over the variable $\vv_{j'}$. If by
hypothesis $\hat{\wv}_{j'}+\hat{\wv}_k\neq\zerov$, this summation
produces $k-2$ terms. Again, we consider each term separately. This
procedure repeats until a subset $\Sc$ of $\{1,\ldots,k\}$ is found,
such that $\sv = \sum_{i\in \Sc} \hat{\wv}_i=\zerov$.

In this case, the contribution of the $n$-th sum is given by
$r-(k-n)$ where $n=|\Sc|$ is the cardinality of $\Sc$.  Overall,
after $n$ sums the total contribution is
    \[ (-1)^{n-1} (n-1)!(r-(k-n))\sum_{\substack{\vv_j, j\in \{1,\ldots,k\}-\Sc
        \\ \neq}} \prod_{j\in
     \{1,\ldots,k\} -\Sc}\zeta^{\vv_j\Tran\hat{\wv}_j}\] The factor $(n-1)!$
    accounts for the number of permutations of the elements in $\Sc$,
    once the first element is fixed (remember that we arbitrarily
    chose the first variable of the summation).  The factor
    $(-1)^{n-1}$ takes into account that we summed $n-1$ times with
    the condition $\hat{\wv}\neq\zerov$, which implies $n-1$ sign
    changes. Eventually,
the term $\sum_{\substack{\vv_j, j\in \{1,\ldots,k\}-\Sc
        \\ \neq}} \prod_{j\in \{1,\ldots,k\}-\Sc}\zeta^{\vv_j\Tran\hat{\wv}_j}$
    is similar to the last term in~(\ref{eq:proof_lemma1_1}) where
    only $k-n$ variables $\vv$ are involved.

   This procedure repeats until we sum over all variables $\vv$.
   This is equivalent to check if for all possible partitions of $\{1,\ldots,k\}$
   in $h$ subsets $\Pc_1,\ldots, \Pc_h$, $h=1,\ldots,k$ the condition
   $\sv_1= \sv_2= \cdots = \sv_h = \zerov$ holds, with $\sv_j = \sum_{i\in
     \Pc_j} \hat{\wv}_i$, $n_j = |\Pc_j|$, and $\sum_j n_j = k$.
   In this case, the contribution is given by
   \[ \prod_{j=1}^h (-1)^{n_j-1} (n_j-1)!
     p_r(n_1,\ldots,n_h) \]
   and it is $0$ otherwise.
Here $p_r(n_1,\ldots,n_h) = (r-(k-n_1))(r-(k-n_1-n2))\cdots (r-(k-n_1-n_2-\cdots-n_{h-1}))$.

   In conclusion, we can write
   \[ \sum_{\substack{\vv_1,\ldots,\vv_k \\ \neq}}
\zeta^{\vv_1\Tran\hat{\wv}_1+ \cdots + \vv_k\Tran\hat{\wv}_k} =
\sum_{h=1}^k  \sum_{\omegav' \in \Omega_{k,h}}
u(\omegav') p_r(\omegav') \prod_{j'=1}^h\delta\left(\sum_{i'\in
     \Pc_{j'}(\omegav')} \hat{\wv}_{i'}(\omegav) \right)
\]
where $u(\omegav')=(-1)^{k-h}\prod_{j'=1}^h (|\Pc_{j'}(\omegav')|-1)!$
and $p_r(\omegav')$ is a polynomial in $r$ of degree $h$.  For large
$r$, $p_r(\omegav') \simeq r^h$, thus proving the lemma.

\end{document}